\documentclass[]{spie}  

\usepackage{amsmath,amsfonts,amssymb}
\usepackage{CJK}
\usepackage{graphicx}
\usepackage[colorlinks=true, allcolors=blue]{hyperref}

\title{Commissioning results from the Robo-AO-2 facility for rapid visible and near-infrared AO imaging}

\author[a]{Christoph Baranec*}
\author[a]{James Ou}
\author[b]{Reed Riddle}
\author[a]{Ruihan Zhang}
\author[a]{Luke Mckay}
\author[a]{Rachel Rampy}
\author[a]{Morgan Bonnet}
\author[a]{Iven Hamilton}
\author[a]{Greg Ching}
\author[a]{Jessica Young}
\author[c]{Maïssa Salama}
\author[a]{Paul Barnes}
\author[a]{Shane Jacobson}
\author[a]{Peter Onaka}
\author[a]{Mark Chun}
\author[a]{Zachary Werber}
\author[$\!$]{Keith Powell}
\author[d]{Marcos A. van Dam}
\author[a]{Benjamin Shappee}

\affil[a]{Institute for Astronomy, University of Hawaii at Manoa, Hilo, HI 96720 USA}
\affil[b]{California Institute of Technology, Pasadena, CA 91125 USA}
\affil[c]{University of California, Santa Cruz, CA 95064, USA}
\affil[d]{Flat Wavefronts, Christchurch 8022, New Zealand}

\authorinfo{* baranec@hawaii.edu; phone 1-808-498-9817; http://robo-ao.org/}

\begin{document} 
\maketitle

\begin{abstract}
We installed the next-generation automated laser adaptive optics system, Robo-AO-2, on the University of Hawaii 2.2-m telescope on Maunakea in 2023. We engineered Robo-AO-2 to deliver robotic, diffraction-limited observations at visible and near-infrared wavelengths in unprecedented numbers. This new instrument takes advantage of upgraded components, manufacturing techniques and control; and includes a parallel reconfigurable natural guide star wavefront sensor with which to explore hybrid wavefront sensing techniques. We present the results of commissioning in 2023 and 2024.
\end{abstract}


\section{INTRODUCTION}
\label{sec:intro}  

   \begin{figure} [ht]
   \begin{center}
   \includegraphics{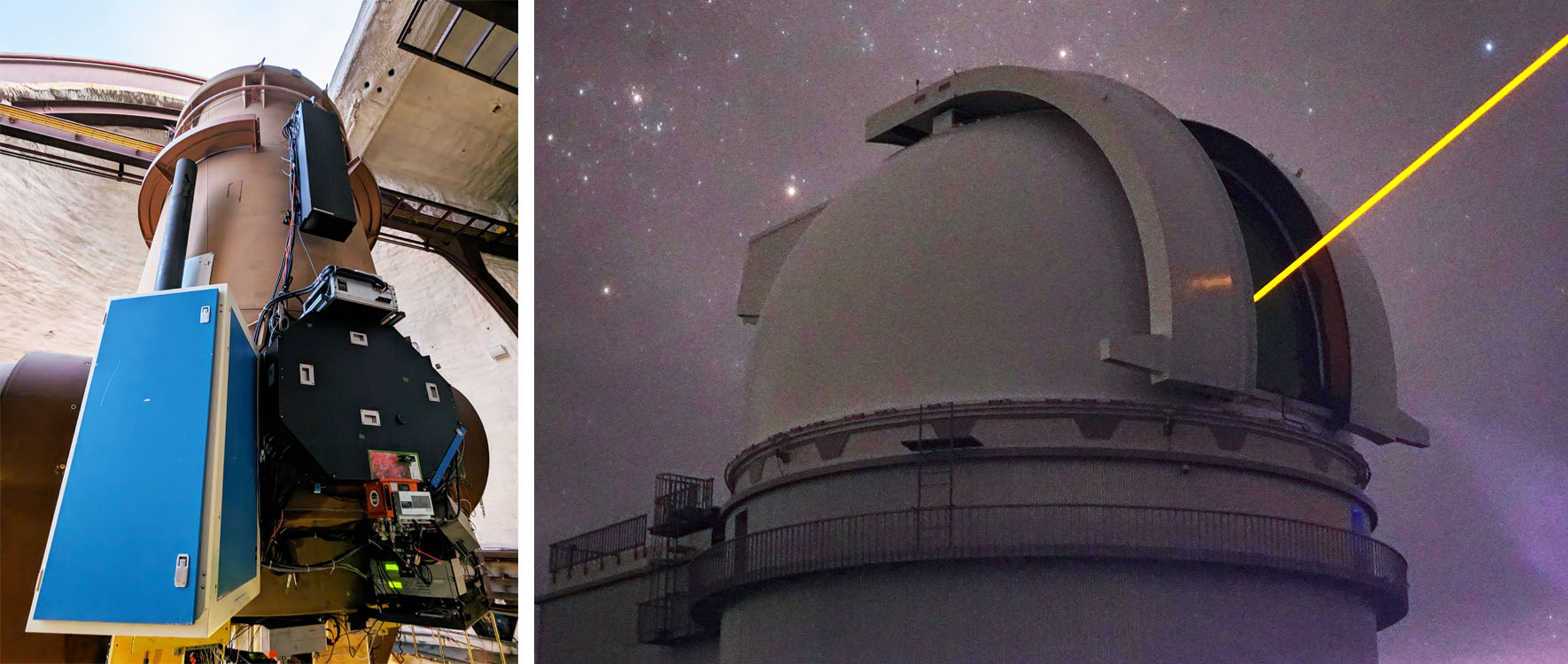}
   \end{center}
   \caption[example] 
   { \label{fig:1} 
\textit{Left:} Robo-AO-2 installed on the North bent-Cassegrain port of the University of Hawaii 2.2-m telescope. \textit{Right:} The 15W UV laser beam projecting from the telescope dome.}
   \end{figure} 

Robo-AO-2 is a robotic laser adaptive optics system built by the Institute for Astronomy for the University of Hawaii 2.2-m telescope (UH2.2m) on Maunakea, Hawaii (see Fig.~\ref{fig:1}). It is based on the prototype Robo-AO system\cite{Baranec2013, Baranec2014, Baranec2021} which was used at the Palomar 1.5-m telescope\cite{cenko} (2011-2015), the 2.1-m telescope at Kitt Peak \cite{RAO_KP} (2015-2018) and then at the UH2.2m \cite{LASSO} (2019-2020). The core innovation with the Robo-AO platform is the full automation of the adaptive optics system, science instruments and data reduction. When paired with a similarly automated telescope, this combination allows for very efficient survey, monitoring, and target-of-opportunity observing. Over its time in service, we used Robo-AO to produce high angular resolution images of over 18,000 unique targets that appear in 54 refereed  publications and that supported 5 Ph. D. dissertations\cite{website}. A non-exhaustive list of scientific investigations that are uniquely enabled by the new Robo-AO-2 system are described in Baranec, et al. 2018\cite{Baranec2018}.

\section{INSTRUMENT DESCRIPTION}
\label{sec:instrument}

The design of Robo-AO-2 follows that of the prototype system with several improvements to the performance and capability. The preliminary design is presented in Baranec, et al. 2018\cite{Baranec2018}, with more specific details highlighted in this section. Robo-AO-2 comprises a UV laser projector, a bent-Cassegrain mounted adaptive optics system with low-noise, high-speed visible and infrared imaging arrays that can double as tip-tilt sensors, a reconfigurable stellar wavefront sensor, and a set of electronics and computers with additional functionality.

\subsection{Optical Baseplate}

The mechanical foundation of the adaptive optics system is an optical baseplate upon which all of the optics, mechanisms, cameras, etc., are mounted. The baseplate is then mounted directly to a fixed steel ring on the telescope. Robo-AO used a 25.4 mm thick square shaped aluminium plate, roughly 1 m on a side, which was pocketed on the underside for weight relief without significantly compromising stiffness. For Robo-AO-2, we required something of similar or greater stiffness, and a baseplate that minimized the effects of thermal contraction as we would be assembling the instrument at our facility in Hilo, $T\sim25\arcdeg$C, and be using the instrument at Maunakea, $-5\arcdeg$C $<T<5\arcdeg$C. Specifically, we would be using low coefficient of thermal expansion (CTE) glass/ceramic materials for the optical relay as opposed to the diamond turned aluminum used with Robo-AO which naturally compensated some of the CTE effects of an aluminium baseplate. The Robo-AO-2 baseplate was designed as an irregular octagon, with an effective footprint of 1041 mm $\times$ 1295 mm.  

   \begin{figure} [ht]
   \begin{center}
   \includegraphics{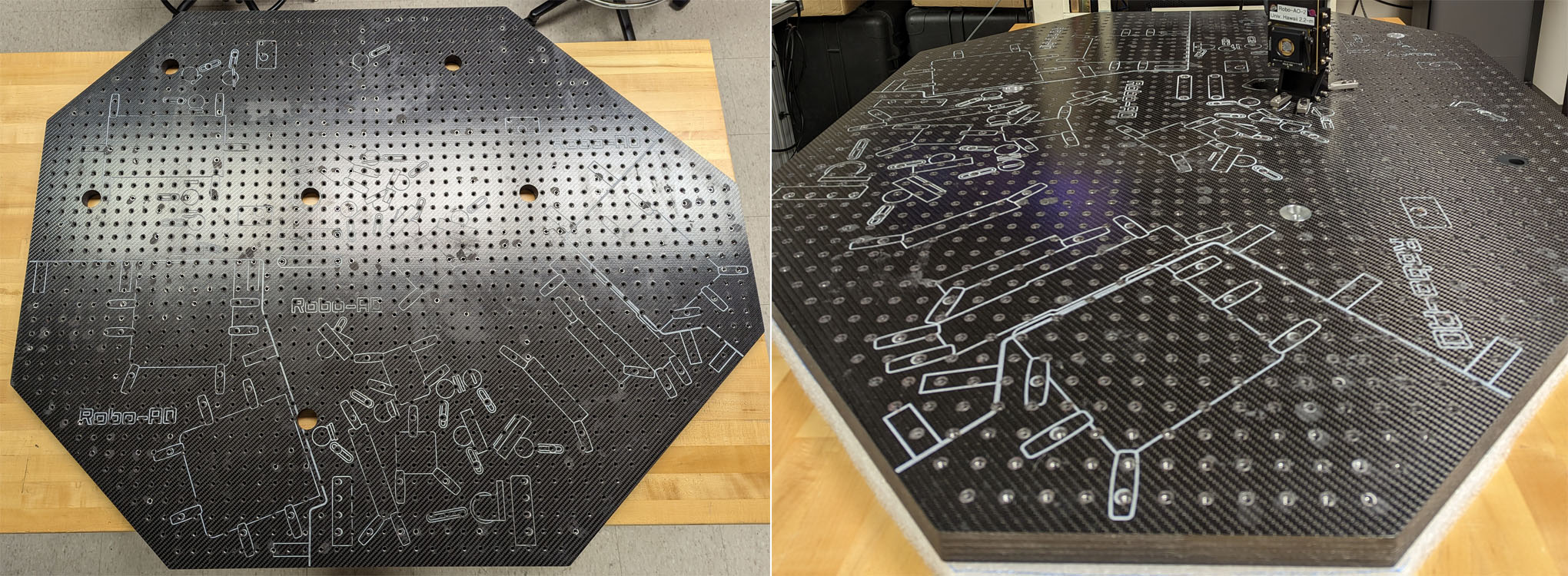}
   \end{center}
   \caption[example] 
   { \label{fig:baseplate} 
The Robo-AO-2 solid carbon fiber optical baseplate. An outline of optical and mechanical components was engraved on the surface to assist with assembly.}
   \end{figure}

In 2021, we purchased a honeycomb carbon fiber baseplate from a vendor that specializes in high precision breadboards, benches and structural parts made of carbon fiber reinforced polymer. The baseplate is 31 mm thick and weighs 31 kg. During the laboratory integration of optics on the baseplate, we noticed that when clamping optical posts, there would be a deflection of neighboring components due to the surface layer of the baseplate bending. This ultimately resulted in our inability to align the natural guide star wavefront sensor as its optical path crossed nearby the optical relay feeding this sensor. Despite this, once the rest of the system was ready, we installed Robo-AO-2 on the UH2.2m in August of 2022. During our off-sky testing, we found that the registration of the deformable mirror with the laser wavefront sensor lenslet array was neither stable nor repeatable when moving the telescope, and that vibrations from the infrared camera Stirling cooler were causing images at the science focal plane to vibrate by over 1\arcsec. On-sky, we were unable to maintain the precise alignment of the laser light required through the Pockels cell, so were unable to close the adaptive optics loop. The deficiencies caused by this initial baseplate led us to seek an alternative.

We purchased a solid carbon fiber optical baseplate from Protech Composites (Vancouver, WA; see Fig.~\ref{fig:baseplate}) after testing the stiffness of samples provided by the vendor. When clamping optical posts in close proximity, we found the component deflections to be smaller than with the original aluminium Robo-AO baseplate, and the resonant frequencies were empirically much higher than the honeycomb baseplate. The new baseplate was delivered at the end of 2022, and is 25.4 mm thick and weighs 44 kg. We proceeded to transfer all of the components to this baseplate in 2023, including the natural guidestar wavefront sensor. We have not yet detected any issues with alignment or vibrations during subsequent on-sky commissioning work.

\subsection{Telescope Simulator, Calibration Optics and Alignment Tools}

To assist with the alignment of the Robo-AO-2 optics, and for later calibration of the adaptive optics system, we employed a telescope simulator subsystem. The telescope simulator is mounted to its own carbon fiber baseplate and attaches to the main optical baseplate through a kinematic mechanism. The telescope simulator employs three different light sources: a multimode octagonal fiber fed by a 5 mW, $\lambda=355$ nm laser with a high-speed fiber agitator to simulate the laser guide star projected to 10 km, a single-mode optical or near-infrared fiber fed by an incandescent source to simulate a star, and a visible LED illuminated pinhole mask that is used to measure the distortion over the full field of the optical relays. The pinhole mask is based on previous work\cite{service} and is a lithographically produced mask with a rectilinear pattern of 100 $\mu$m diameter holes every 500 $\mu$m that cover beyond the maximum unvignetted $72\arcsec$ diameter science field of view. The central and two nearby holes are oversized to 150 $\mu$m to aid in field, coordinate and parity registration. The single-mode fiber and pinhole mask are interchangeable in the optical train with the use of a remotely actuated fold mirror. The visible sources are collimated and the UV source is almost collimated (to account for the difference in field) before being combined with an uncoated beam-splitter cube and passed through a common stop to simulate the telescope primary stop. The stop can be manually swapped between a circular aperture and an annular aperture which simulates the secondary obscuration and spiders. The sources are then refocused with a spherical mirror and another uncoated beam-splitter cube is used to direct the light towards the adaptive optics relay. This optical path crosses the location where the natural telescope focus would land, and a fold mirror on a linear stage is placed at this location so one can choose whether to feed the adaptive optics system with either light from the telescope simulator or the telescope. The measured residual wavefront error coming from the telescope simulator (with wavefront curvature removed from the F/10 beams) was 43 nm RMS from the visible light source and 24 nm RMS from the UV source.  

   \begin{figure} [ht]
   \begin{center}
   \includegraphics[height=5cm]{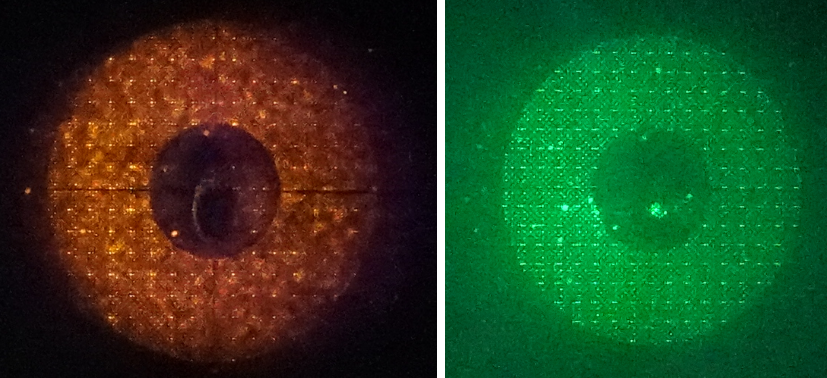}
   \end{center}
   \caption[example] 
   { \label{fig:footprints} 
Scattered light images of the UV (left) and visible (right) beams on the deformable mirror surface.}
   \end{figure}

To perform the optical alignment of the adaptive optics system, we used many typical tools such as fiducial targets and shear plates, but there were three of particular mention:

1. We used a large format CMOS sensor with $3.5 \mu$m pixels on a linear stage to accurately measure the z-location of focal points and to measure beam sizes directly.

2. We used a Sony $\alpha$7S II DSLR camera modified to be sensitive to $\lambda=355$ nm, combined with a Coastal Optics UV-VIS-IR 60mm Macro lens, to measure beam sizes and location using scattered light off of reflective optics. This was particularly helpful when overlapping the UV and visible beam footprints on the deformable mirror surface to ensure that both the registration and z-distance was correct (e.g., see Fig.~\ref{fig:footprints}). 

3. We used Imagine Optic HASO4 Broadband and HASO3-128GE2 Shack-Hartmann wavefront sensors for the final alignment of optics. These were especially useful during the alignment of the off-axis parabolic (OAP) mirrors, both in collimated space (after OAPs 1 and 3) and in converging space (after OAPs 2 and 4). The HASO4 Broadband is additionally sensitive to $\lambda=355$ nm, so we could measure the wavefront error of the collimated beam just before the lenslet array in the laser wavefront sensor which was measured to be 22 nm RMS.

\subsection{Optical Layout}

\begin{figure} [ht]
   \begin{center}
   \includegraphics[width=6in]{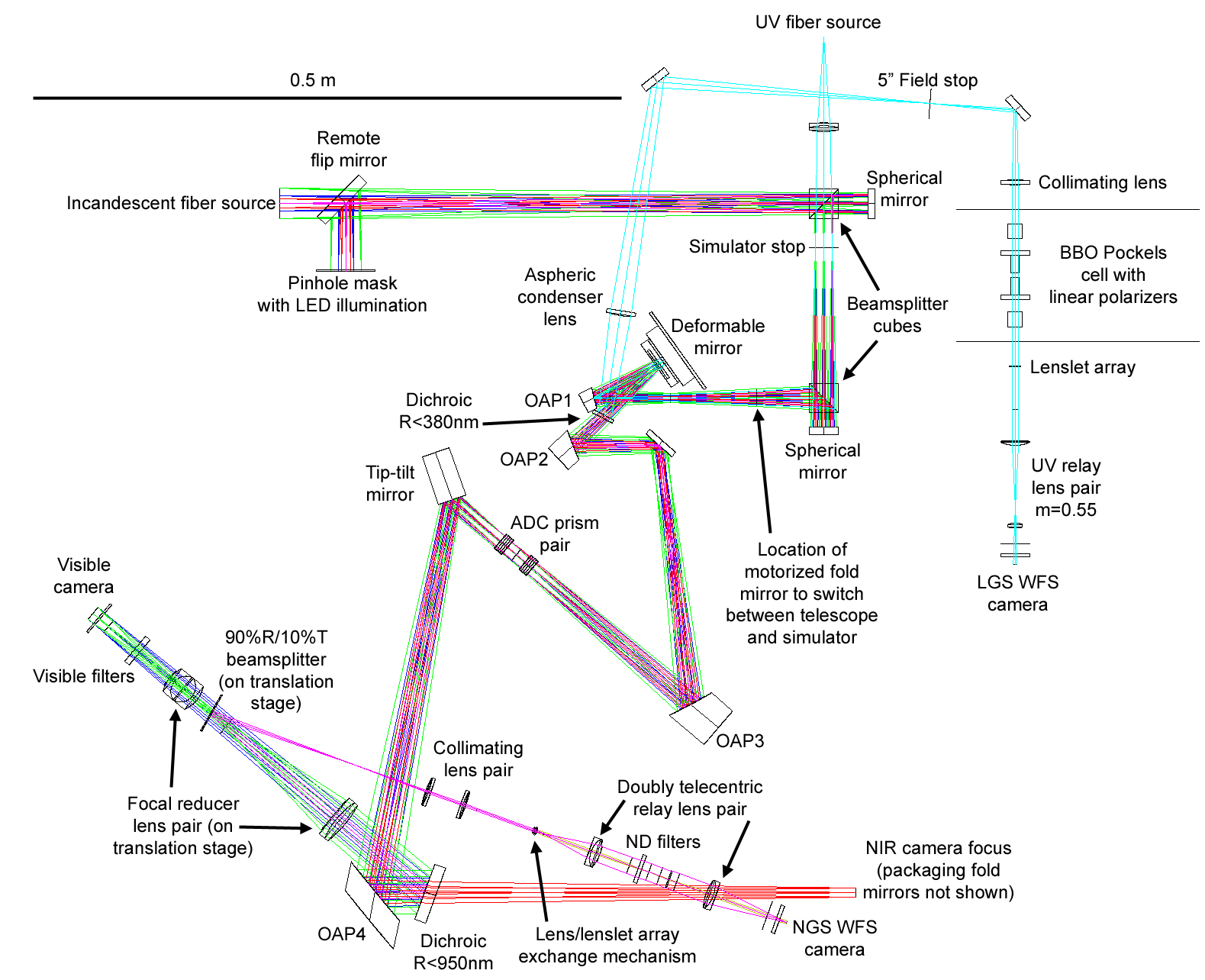}
   \end{center}
   \caption[example] 
   { \label{fig:zemax} 
Optical model of Robo-AO-2.}
   \end{figure}

\begin{figure} [ht]
   \begin{center}
   \includegraphics{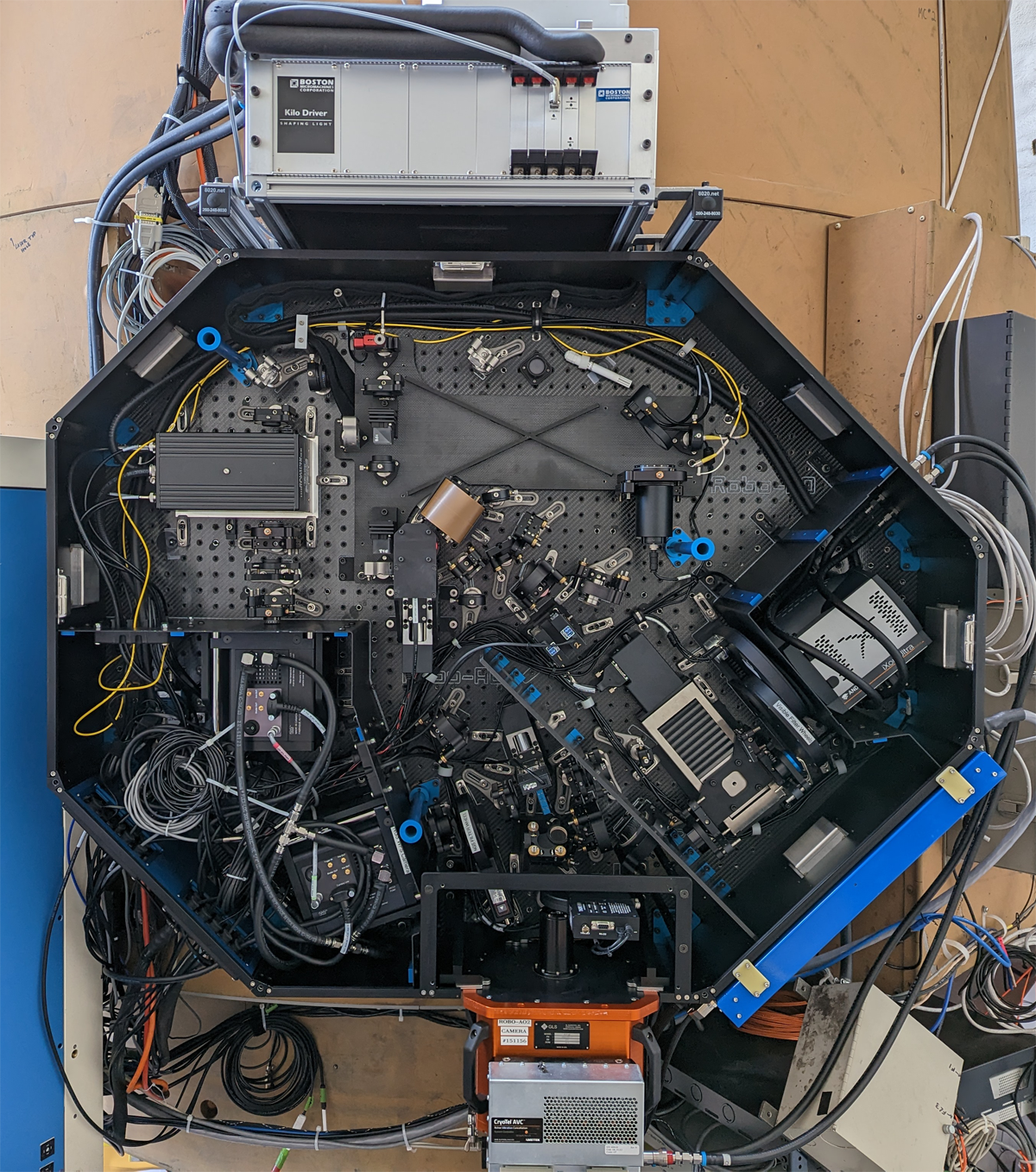}
   \end{center}
   \caption[example] 
   { \label{fig:inside} 
Photograph of the Robo-AO-2 adaptive optics system on the UH2.2-m with the cover removed.}
   \end{figure}

The Robo-AO-2 optical system was designed using the commercial software package Zemax 13 Release 2 SP6 64-bit. Subsystems and components were designed separately, then merged together with six configurations to create the final model. A layout of this model is shown in Fig.~\ref{fig:zemax} and a photo of the instrument with the cover taken off appears in Fig.~\ref{fig:inside}.

The telescope simulator mimics the focal ratio and pupil position of the UH2.2m telescope. From the location of the motorized fold mirror, light from either the simulator or the telescope propagates to OAP1. Stellar light is collimated (UV light is slightly diverging) and forms an image of the telescope pupil on the deformable mirror spanning 17 actuators. The deformable mirror is a Boston Micromachines DM492-3.5 with a MgF2 window with an AR coating effective from the near-UV to the near-IR. This mirror has been calibrated for open-loop operation (based on work by \citeonline{openloop}) which allows us to command the desired surface shape as opposed to actuator voltages, but this has not yet been implemented in our software. Light is reflected off of the deformable mirror surface toward a dichroic mirror that reflects light with a wavelength of less than 380 nm.

The reflected UV light is refocused with a low-power aspheric condenser lens. At the 10km focus is a 5\farcs0 field stop. A slight decentering of this lens mitigates most of the optical error coming off of OAP1. The light is then collimated to a 5.6 mm diameter beam which passes through a double crystal Pockels cell ($\beta$-BaB$_2$O$_4$; 7mm$\times$7mm$\times15$mm) with crossed polarizers at the entrance and exit of the cell. The entrance polarizer is matched to the returning polarization of the laser beacon which is linearly polarized in the East-West direction on sky. A lenslet array is located at the reimaged pupil just after the Pockels Cell, has a pitch of 350$\mu$m, and samples the beam with 16 lenslets (Fried geometry). The resulting Shack-Hartmann image plane is then reimaged on to the laser guide star wavefront sensor camera which is a Nüvü Camēras HNü 128 with an astro processed Teledyne e2v CCD60 detector. The measured quantum efficiency of this specific detector is 68\% at $\lambda$ = 355 nm. 

The visible and near-infrared light pass through the dichroic mirror and are refocused by OAP2. The intermediate focus is collimated by OAP3 which creates a 12 mm diameter image of the pupil at the mid-point of an atmospheric dispersion correcting (ADC) prism pair. Immediately after the ADC is a piezo driven tip-tilt mirror. Next is the final OAP4 which formats the light to a telecentric F/35 focus. Immediately after OAP4 is a dichroic that reflects light of $\lambda<950$ nm to the visible camera. Near-infrared light transmitted through the dichroic is directed by a pair of mirrors through a warm filter wheel to the custom near-infrared camera.  

All OAPs are made of Corning ULE 7972 and are figured and polished to a surface of better than 32 nm RMS over their clear aperture (which exceeds the reflected beam footprints). Both OAP1 and the fold mirror that directs light from the telescope into the instrument are coated with a special enhanced silver coating from Infinite Optics that has $>$99\% reflectivity at the laser wavelength while maintaining the benefits of high reflectivity from silver over visible and near infrared wavelengths. OAPs 2, 3 and 4, along with the fold mirror immediately after OAP2 are coated with a standard enhanced silver coating from L \& L Optical Coatings. Using a silver coating on optics leading to the science cameras helps to attenuate any laser light that leaks through the short wavelength dichroic mirror.

\begin{figure} [ht]
   \begin{center}
   \includegraphics[height=2.5in]{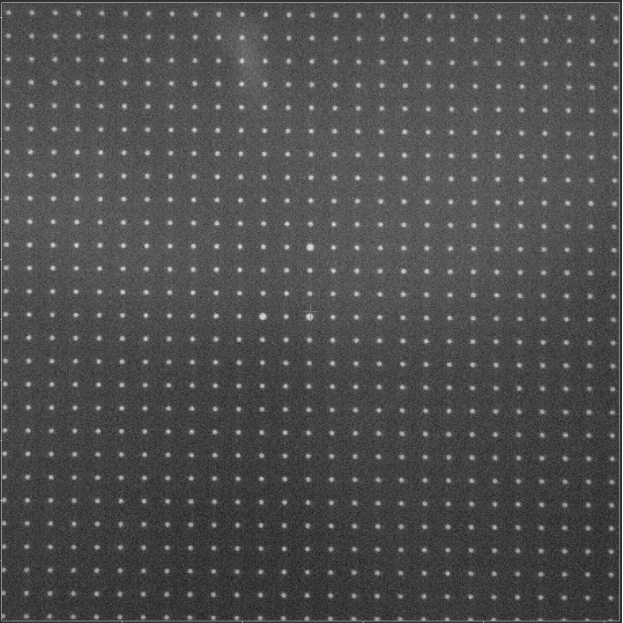}
   \end{center}
   \caption[example] 
   { \label{fig:distortion} 
Image of the astrometric calibration mask on the visible science camera.}
   \end{figure}

During the design of the OAP relays, we paid particular attention to optical distortion of the science field. The main contribution to optical distortion comes from the OAP1-OAP2 pair where the angle of the incidence of the on-axis ray to each OAP is $14\arcdeg$ and $21.5\arcdeg$ respectively, constrained due to passing incoming light around the packaging of the deformable mirror and having the intermediate focus not interfere with other system components.  At the intermediate focus of the OAP1-OAP2 relay, the maximum optical distortion over the $36\arcsec$  square visible science field is 2.0\%. With careful optimization of the OAP3-OAP4 pair, the designed maximum optical distortion of the final image plane is reduced to 0.03\%. We confirmed the very low distortion of the as-built system using the LED illuminated pinhole mask in the telescope simulator (see Fig.~\ref{fig:distortion}).

\subsection{Atmospheric Dispersion Corrector}

The atmospheric dispersion corrector prism pair is designed to minimize chromatic aberration caused by the Earth's atmosphere. Each prism is made of two glasses, S-TIM3 and N-KZFS8, that are optically bonded together and with broadband anti-reflection coatings on the exposed surfaces. The two glasses are different in their CTE by only $1\times10^{-7}/$C$\arcdeg$ which minimizes thermally induced stress on the bonded pair. The ADC is designed to work from $\lambda=420$ nm to 1.65 $\mu$m and up to a Zenith angle of 60\arcdeg.

During the design of the ADC, we were careful to minimize downstream pupil shifts and anamorphic distortion which is maximized when the prisms are clocked at 90$\arcdeg$ with respect to each other (for this design it occurs at a Zenith angle of $50\arcdeg$). At this clocking, there is a maximum of 0.2\% distortion at the corner of the 36$\arcsec$ square field, and the pupil is distorted by a maximum of 16\% of a subaperture around the edge of the pupil on the stellar wavefront sensor in the 16 across sampling mode. These errors can be reduced by restricting the maximum Zenith angle of correction in the original design. When making precision astrometry measurements near high Zenith angles, we intend to use the telescope simulator pinhole mask to calibrate the actual distortion caused by the instrument. We intend to use the stellar wavefront sensor much closer to Zenith, so the residual pupil distortion is acceptable. 

The tolerances on the fabrication of the ADC were tighter than many of our other optics. Specifically, we found that the typical manufacturer tolerance on index of refraction and dispersion were insufficient to meet our requirements. During the fabrication of the ADC, we obtained the melt data for each of the two glasses and then re-optimized the design angles on the prisms before having them manufactured. 

\subsection{Imaging Cameras}

Robo-AO-2 employs three different imaging cameras (see Tab.~\ref{tab:cameras}), all co-focal with each other. 

\subsubsection{Pointing Camera}

We use a ZWO ASI1600MM camera with a Panasonic MN34230ALJ sensor as a wide-field pointing camera. The camera is primarily used during the initial setup of Robo-AO-2 on the UH2.2-m to find the pointing of the instrument on sky. It is unfiltered for maximum sensitivity, which also allows us to image the defocused laser beam on sky - aiding in the fine alignment of the laser projector optical axis to that of the telescope. This camera is accessed by moving the motorized fold mirror that switches between the telescope simulator and telescope; when the telescope simulator is feeding the adaptive optics system, the motorized fold mirror is out of the way of the pointing camera.  

\begin{table}[ht]
\caption{Robo-AO-2 imaging cameras} 
\label{tab:cameras}
\begin{center}       
\begin{tabular}{|c|c|c|c|c|} 
\hline
\rule[-1ex]{0pt}{3.5ex}  Detector & Technology & Format & Pixel Scale & Field  \\
\hline
\rule[-1ex]{0pt}{3.5ex}  MN34230ALJ & CMOS & $4656\times3520$ & 0\farcs035 & $2\farcm7\times2\farcm0$ - seeing limited\\
\hline
\rule[-1ex]{0pt}{3.5ex}  CCD201-20 & EMCCD & $1024\times1024$ & 0\farcs035 & $36\arcsec\times36\arcsec$ \\
\hline
\rule[-1ex]{0pt}{3.5ex}  CCD201-20 & EMCCD & $1024\times1024$ & 0\farcs070 & $72\arcsec$ dia. unvignetted with focal reducer \\
\hline
\rule[-1ex]{0pt}{3.5ex}  SAPHIRA & LmAPD & $320\times256$ & 0\farcs065 & $21\arcsec\times17\arcsec$ \\
\hline 
\end{tabular}
\end{center}
\end{table}

\subsubsection{Visible Camera}

Our visible science camera is an Andor iXon Ultra 888 camera with a Teledyne e2v CCD201-20 EMCCD detector. Immediately in front of the camera are two 8-position filter wheels that are populated with a set of Asahi Spectra all-dielectric SDSS filters as well as various narrow-band and long-pass visible filters. Before the filter wheels is also a linear motion stage with three selectable positions: 1. no optics - the F/35 beam from OAP4 is imaged directly on the detector; 2. a focal reducer lens pair which effectively doubles the field of the view of the visible camera while maintaining the same final focus position; and 3. a 1 mm thin beam splitting mirror that reflects 90\% of light to the natural guide star wavefront sensor and transmits 10\% of light to the visible science camera.  

\subsubsection{Near-infrared Camera}

\begin{figure} [ht]
   \begin{center}
   \includegraphics{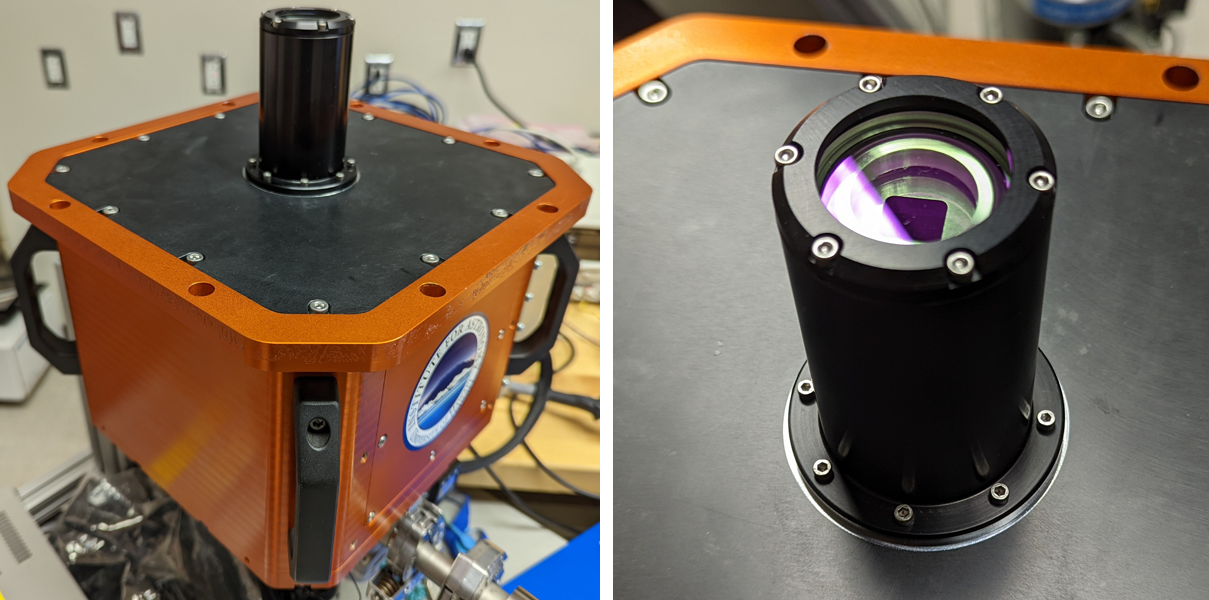}
   \end{center}
   \caption[example] 
   { \label{fig:ircamera} 
The Robo-AO-2 near-infrared camera cryostat showing the cold snout and entrance window.}
   \end{figure}

We reused Robo-AO's Hawaii Aerospace Stirling cooler cryostat\cite{raoirc, LASSO} as the near-infrared camera on Robo-AO-2 with some improvements (see Fig.~\ref{fig:ircamera}). The new detector is a linear-mode avalanche photodiode array (LmAPD): a Mark 20 Selex ES Advanced Photodiode for High-speed Infrared Array (SAPHIRA) with an ME-1001 Readout Integrated Circuit, similar to that used with Keck observatory's near-infrared pyramid wavefront sensor\cite{keckpyramid}. We use the single-board PB1 ‘PizzaBox’ readout electronics that were developed at the Institute for Astronomy to readout the detector. These electronics use 32 readout channels, each capable of a 2 Mpixel/sec sampling rate, for a maximum full-frame read rate of $\sim$800Hz. We employ the read-reset mode of the readout electronics to achieve the maximum full well for every exposure.

The default cryostat configuration includes a 4 inch diameter entrance window which is appropriate when testing HAWAII-4RG detectors, but puts unnecessary thermal loading on the cryostat's outer radiation shield. To address this issue, we built our own custom front end that restricts the incoming focal ratio to match that provided by OAP4, and have extended the radiation shield by 3 inches with a cold snout. The cold snout reduces the solid angle seen by the detector by a factor of 10. At the base of the cold snout is a thermal blocking filter that blocks light longward of $\lambda=1.8\mu$m. Near the thermal blocking filter we have attached extra getter material to prevent volatiles freezing on the filter as we have occasionally seen in early iterations of this camera. Immediately before the entrance window of the camera, we have a warm six-position filter wheel with Asahi Spectra Mauna Kea Observatories Y, J and H filters, along with a clear and blocking filter.

\subsection{Natural Guide Star Wavefront Sensor}

\begin{figure} [ht]
   \begin{center}
   \includegraphics{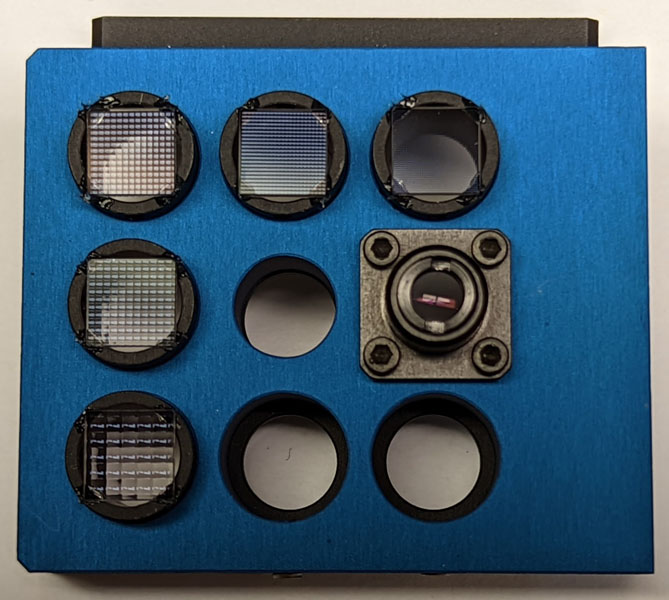}
   \end{center}
   \caption[example] 
   { \label{fig:lenslet} 
The natural guidestar wavefront sensor lenslet arrary exchange holder.}
   \end{figure}

The natural guide star wavefront sensor in Robo-AO-2 is primarily intended to support demonstrations of hybrid laser-stellar wavefront sensing\cite{Baranec2018, hapa, pulsemk}. This requires that the sensor have adjustable spatial and temporal sampling of the wavefront. To mitigate risk, we designed a Shack-Hartmann wavefront sensor that has multiple different selectable lenslet arrays via an x-y stage, similar to the sensor employed on the PALM-3000 adaptive optics system\cite{P3K_WFS}. We have five different lenslet arrays that are available that span either 16, 8, 5, 4 or 2 subapertures across the reimaged 2.4 mm diameter pupil. The radius of curvature of every lenslet array is the same. In addition to the lenslet arrays in the exchange mechanism (see Fig.~\ref{fig:lenslet}), there is a passthrough hole in the center for pupil imaging, a lens which allows the sensor to be used as a tip-tilt sensor (co-focal with the lenslet arrays), and two empty positions for future lenslet arrays.

A doubly telecentric lens pair is used to reimage the lenslet array focal plane on a Nüvü Camēras HNü 128 camera with an astro processed Teledyne e2v CCD60 detector. In between these two lenses is a filter wheel with reflective neutral density filters to allow us to explore different brightness guide stars without needing to repoint the telescope to a new target (neglecting the change in sky brightness). Similar to the laser guide star wavefront sensor, there is a 5\farcs0 field stop at the beginning of the wavefront sensor which prevents light from one subaperture from landing in an adjacent subaperture. 

\subsection{Robotic software}

The operational and robotic software for Robo-AO-2 is functionally similar to, and based off of, the software developed for Robo-AO\cite{reed, Baranec2021}. A more thorough discussion of the updates and improvements can be found in these proceedings\cite{software}.

\section{Robotic UH 2.2-m Telescope (Robo88)}
\label{sec:sections}

To maximize the scientific returns from Robo-AO-2, it needs to be installed on a fully automated host telescope. In 2019, we embarked on the Robo88 project to fulfill this requirement (PI: B. Shappee). Robo88 will be an agile robotic observatory with three instruments: Robo-AO-2; STACam, a 10K by 10K wide field-of-view imaging camera ($14\arcmin \times 14\arcmin$); and, the Supernova Integral Field Spectrograph (SNIFS\cite{snifs, scat}), a fully-integrated high-throughput instrument delivering 1\% spectrophotometric observations for point sources even on structured backgrounds. Robo88 will be a powerful new tool for large surveys, rapid transient classification, adaptive optics follow up of exoplanet hosts, and multimessenger astronomy.

\begin{figure} [ht]
   \begin{center}
   \includegraphics[width=4in]{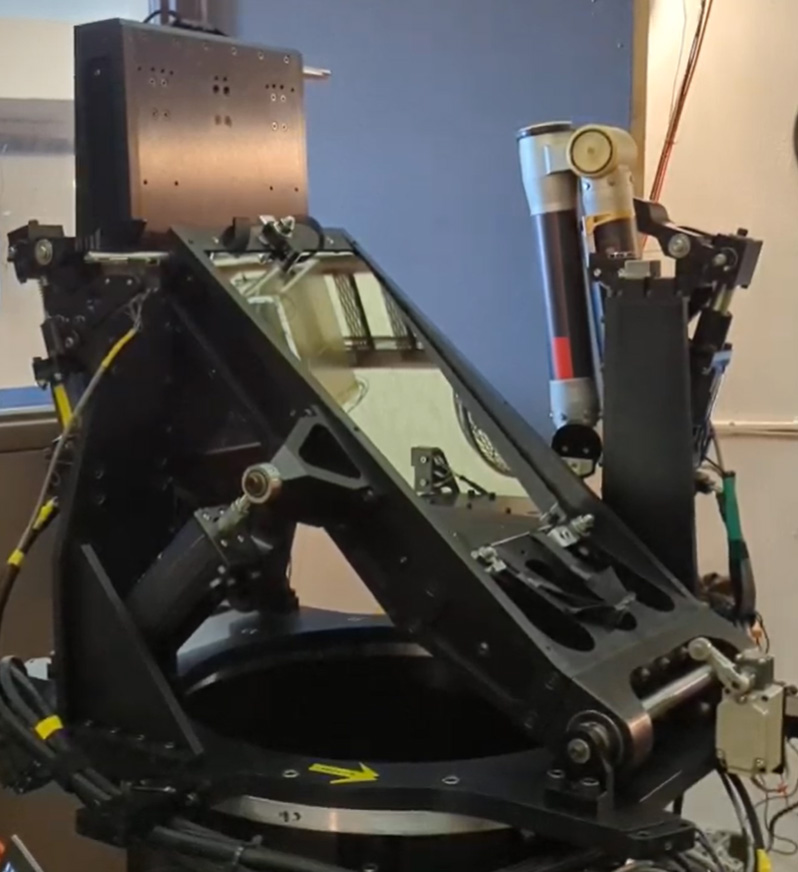}
   \end{center}
   \caption[example] 
   { \label{fig:tertiary} 
The mechanized tertiary mirror system just before being installed at the UH2.2m telescope.}
   \end{figure}

In a robotic observatory, efficiency is key. We are engaged in multiple upgrades to the existing UH2.2-m telescope that will enable and enhance robotic operations. In April 2022 we installed a mechanized tertiary mirror system which allows the telescope to feed any of the three instruments at any time (see Fig.~\ref{fig:tertiary}). The system uses two separately actuated mirrors; they can both be in the stow position which allows light from the secondary to go to the through Cassegrain focus, or one of the two mirrors can be moved into the beam to direct light to either the North or South bent Cassegrain instrument ports. Initial testing shows we can switch between the North and South positions, which requires stowing one mirror before deploying the other, in less than 40 s. We will investigate running the tertiary mechanism faster if driven by future science requirements. The tertiary mirrors supplied by Asahi Spectra are made of fused silica with dimensions of 356 mm$\times$ 229 mm$\times40$mm (thick). The mirrors have a multi-layer dielectric stack reflective coating with an average reflectance of 98.9\% from $\lambda=320$ nm to 1.85 $\mu$m. A stress relieving coating was deposited on the reverse side. We selected a dielectric coating for its robustness and ease of cleaning in place; we do not expect to ever remove these mirrors for service. The measured elliptical clear aperture surface error after coating for both mirrors is less than 28nm RMS and 128nm P-V.

Other improvements to the telescope are at various levels of completion: \begin{itemize}
  \item Replacement of the right ascension and declination drive preload systems and addition of absolute encoders to improve acquisition and tracking of the telescope, and reduce wind-driven tube shake and drive oscillation.
  \item Updating the cable management system.
  \item Added an improved guider and filter wheel to STACam.
  \item Updating the SNIFS control computers.
  \item Development of a robotic telescope control system based on previous work at the Kitt Peak 2.1-m telescope\cite{RAO_KP}, expected mid-2024.
  \item Development of a flexible telescope scheduler.
  \item Realuminizing the telescope primary and secondary mirrors, expected second-half 2024.  
  \item Development a user-friendly pipeline and data archive.
  \item Increasing the robustness and efficacy of the telescope cooling system.
\end{itemize}

\section{Commissioning results}

Robo-AO-2, with the solid carbon fiber optical baseplate, was installed on the UH2.2m in May of 2023. We have been commissioning ever since. As the work on the robotic telescope control system is ongoing, which
includes integration of the absolute telescope encoders, we have been operating Robo-AO-2 in a manual fashion. We have also been troubleshooting adaptive optics performance issues (photoreturn, bandwidth, etc.) that will be discussed in Sec.~\ref{sec:future_work}.

\begin{figure} [ht]
   \begin{center}
   \includegraphics[width=6.5in]{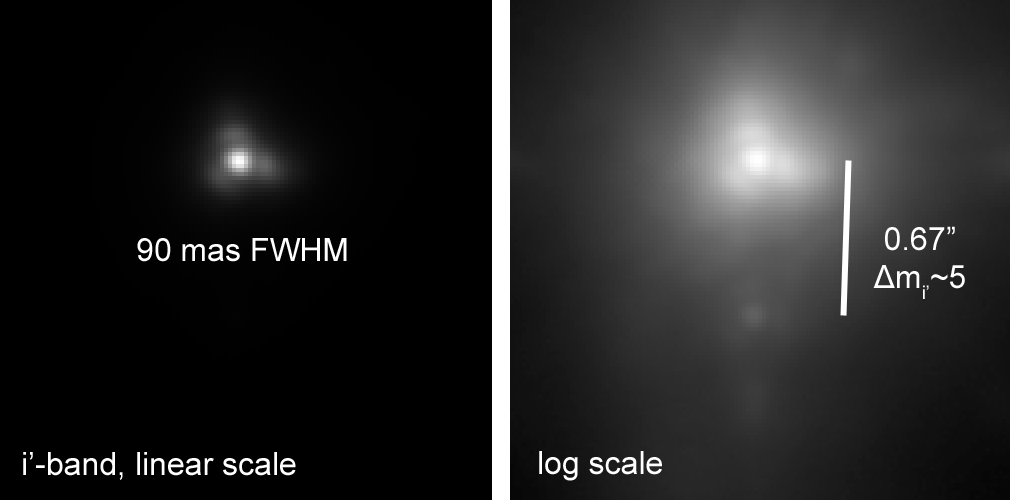}
   \end{center}
   \caption[example] 
   { \label{fig:iband} 
An image of a star in i'-band taken with the Robo-AO-2 visible camera.}
   \end{figure}

\begin{figure} [ht]
   \begin{center}
   \includegraphics[width=6.5in]{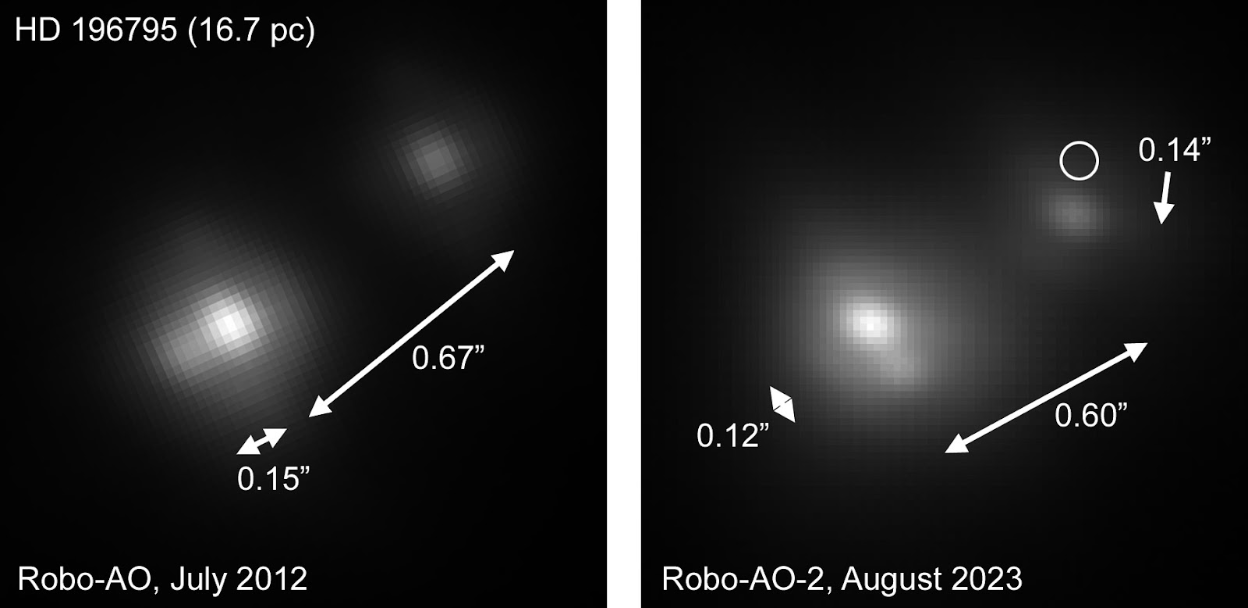}
   \end{center}
   \caption[example] 
   { \label{fig:triple} 
Images of HD 196795 taken over a decade apart with the Robo-AO (i'-band) and Robo-AO-2 (z'-band) systems shown on the same scale and orientation. This demonstrates that Robo-AO-2 can be used to monitor orbits of discovered close binaries.}
   \end{figure}

We have imaged several field stars as well as targets from our previous Robo-AO surveys. All data shown are taken with a series of fast frame rate readouts, typically 0.1s, with post-facto shift and add processing to remove stellar displacement to simulate a long exposure image. Fig.~\ref{fig:iband} shows a 5 minute exposure of a star taken with the visible camera. The FWHM of the point-spread function of the primary star was measured to be 0\farcs09. A candidate stellar companion was measured to be 0\farcs67 away at a contrast of approximately 5 magnitudes. Fig.~\ref{fig:triple} shows a two-minute re-observation of the HD 196795 multiple system that we observed with Robo-AO at Palomar as part of a survey to observe all stars within 25 pc with a declination $> -13\arcdeg$\cite{robo25}. Because the primary is at a distance of 16.7 pc, we would expect to easily see the effects of proper motion over the course of a decade. Further observations will be necessary to discern if these stars are physically associated. 

\begin{figure} [ht]
   \begin{center}
   \includegraphics[width=6.5in]{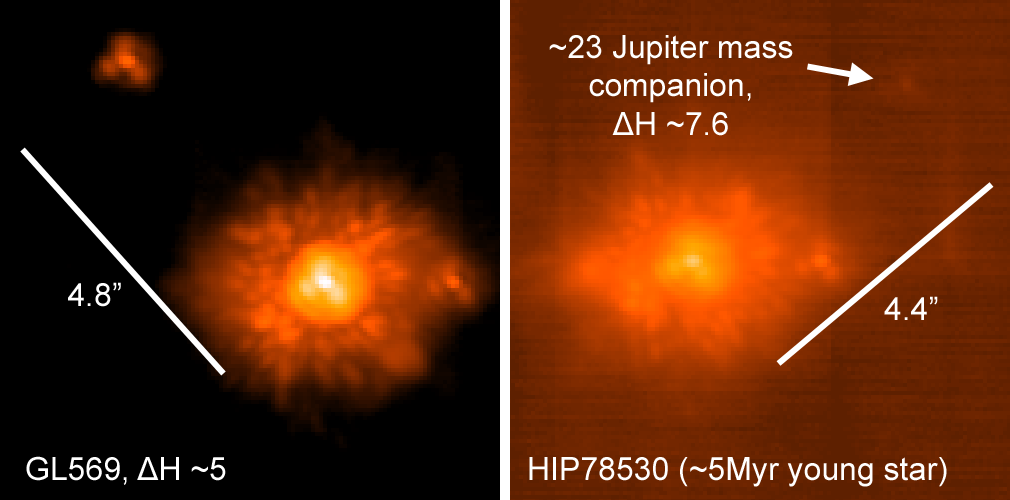}
   \end{center}
   \caption[example] 
   { \label{fig:irdata} 
Images from the Robo-AO-2 near-infrared camera. H-band, log stretch.}
   \end{figure}

\begin{figure} [ht]
   \begin{center}
   \includegraphics[width=3.25in]{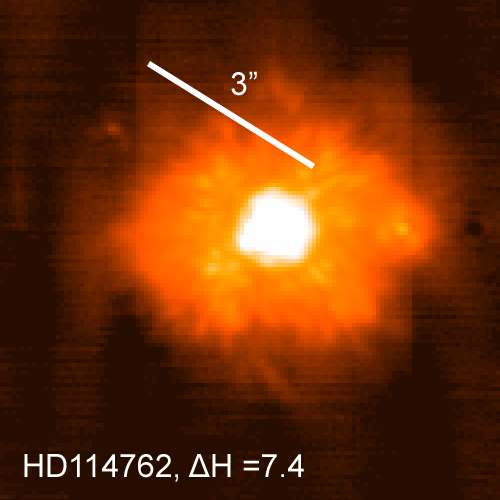}
   \end{center}
   \caption[example] 
   { \label{fig:irdata2} 
Image from the Robo-AO-2 near-infrared camera. H-band, log stretch.}
   \end{figure}

We have also imaged several stars with low mass companions with our near-infrared camera. Appearing in Fig.~\ref{fig:irdata} are GL569\cite{GL569} and HIP78530\cite{HIP78530}, and in Fig.~\ref{fig:irdata2} is HD114762\cite{HD114762}. Images of the stars represent 5 minutes of exposure time in H-band. There is ghost image appearing exactly 32 pixels to the right of each star (corresponding to the 32 readout channels and equivalent to 2\farcs1) which we have confirmed is not optical in nature. There are two analog signal filters on the PB1 readout electronics and the ghost will appear brighter than the source if we use the 284 kHz low-pass filter, however the 7.2 MHz low-pass filter used here results in the ghost remaining (albeit fainter) and higher readnoise. We are continuing to investigate solutions to this issue.   

\section{FUTURE WORK}
\label{sec:future_work}

\subsection{Laser photon return}

One limitation to the current performance of the adaptive optics system is the laser signal photoreturn. The UH2.2m telescope primary and secondary mirrors have unfortunately not been realuminized since 2007 due to various circumstances. We planned to have them recoated in summer of 2023, but this was deferred to the second half of 2024. In the meantime, we have measured the return flux and signal-to-noise ratio (SNR) during commissioning and have found it to be insufficient to meet our performance goals.

When designing the basic parameters of Robo-AO-2 with our error budget spreadsheet (see \citeonline{Baranec2018}), we assumed a reflectivity of the telescope primary and secondary to be 75\% each at the laser wavelength, or 56\% combined. This was generally consistent with our experiences at the Palomar 1.5-m, and Kitt Peak 2.1-m telescopes that are recoated on a regular basis; when the Robo-AO system operated on those telescopes, we measured a SNR range of 6 to 10. For a beacon at a distance of 10 km with a gate depth of 600 m, and running the laser wavefront sensor at 1.5 kHz, we would expect a return of 102 ph-e$^-$/subaperture/exposure which would result in a SNR of 7.1 on the UH2.2m with the total noise dominated by photon noise and the excess noise factor of the EMCCD.

During our on-sky commissioning, we measured the photoreturn which ranged from 55 ph-e$^-$/subap./exp. (SNR=5.1) in July 2023 to 44 ph-e$^-$/subap./exp. (SNR=4.7) in April of 2024. Using these values, we estimate the combined reflectivity of the primary and secondary mirrors at the laser wavelength to be 30\% and 24\%, respectively, or roughly half of what we expect. We borrowed a Surface Optics 410-Solar reflectometer to directly measure the specular reflectivity of the primary, secondary and tertiary mirrors in July 2023, and measured 65\%, 72\% and 94\% respectively in the near-UV ($\lambda=335-380$ nm). This would give a combined reflectivity of 44\%, higher than the 30\% we were estimating from the July 2023 measurements. We note that we measured the new tertiary mirrors in April of 2022 and instead of measuring a reflectivity of 99\%, we measured 105\% in the near-UV (a calibration error) and 99\% over the $\lambda=700-1700$ nm range. If we calibrate our measurements of the telescope mirrors with this anomaly, the combined measured reflectivity would be $44\%\times1.06^{-3}=37\%$, which still leaves a discrepancy of $37\%/30\%$, or 23\%.

We are therefore waiting for the realuminization of the telescope primary and secondary mirrors later this year in anticipation of a laser photoreturn closer to that anticipated in our original error budget calculation. It has been 17 years since the last recoat at the UH2.2m, and we are taking our mirrors to the Canada–France–Hawaii Telescope facility for recoating for the first time. Our telescope staff and engineers are learning this process and we expect to make this a regular occurrence consistent with other major observatories. We will also be instituting regular mirror cleaning and control measures to minimize the amount of contamination that accumulates on the primary and tertiary mirrors (which is why the tertiary mirror reflectivity dropped by 9\% in just one year). We have also just recently acquired our own 410-Solar reflectometer, so we will be tracking the changes in reflectivity more frequently over time and will adjust our maintenance schedule accordingly. 

\subsection{Correction bandwidth}

Another limitation to the current performance of the adaptive optics system is the bandwidth of correction. During commissioning, we compared the power spectral density of the reconstructed disk-harmonic modes measured by our laser wavefront sensor during open- and closed-loop operation. Initially we were using an integral only control law and we measured the 0-dB correction bandwidth to be approximately 80 Hz, despite operating the adaptive optics system at a loop rate of up to 1.8 kHz. For reference, Robo-AO had a loop rate of 1.2kHz with an effective wavefront control bandwidth of 90–100 Hz.

\begin{figure} [ht]
   \begin{center}
   \includegraphics[width=6.5in]{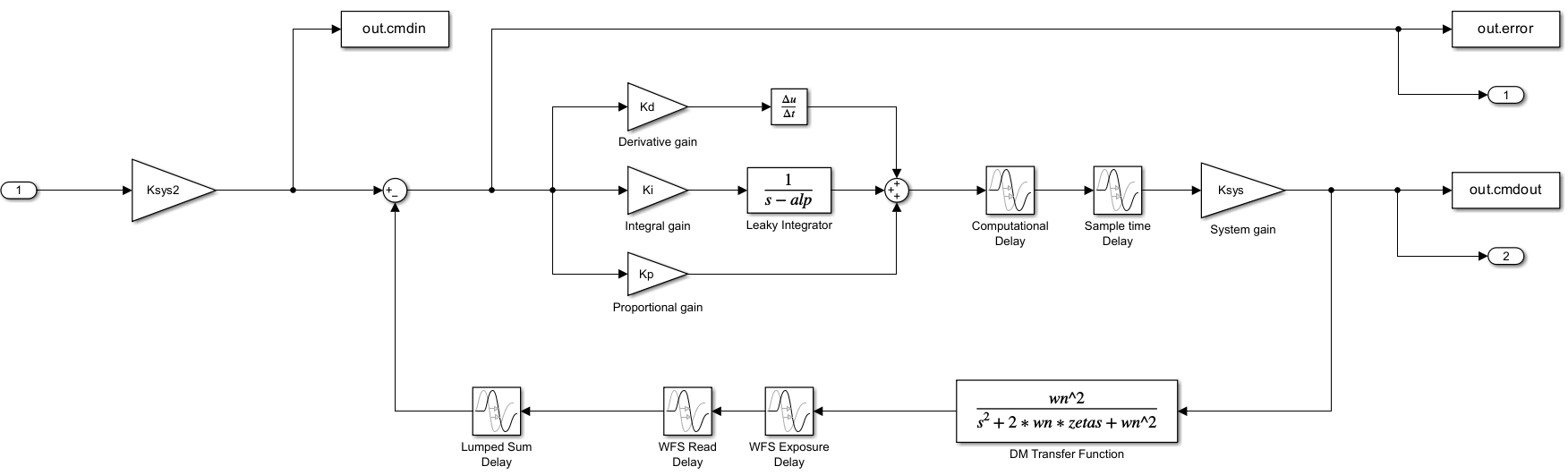}
   \end{center}
   \caption[example] 
   { \label{fig:simulink} 
MATLAB Simulink model of the Robo-AO-2 adaptive optics control loop.}
   \end{figure}

To investigate this apparent inconsistency, we created a simple model of the control system using the Simulink toolbox in MATLAB (Fig.~\ref{fig:simulink}) based on the previous work of one of us\cite{keith_phd}. We validated the model against open- and closed-loop operation of the adaptive optics system running on the internal telescope simulator (where the SNR $>10$ and disturbances are caused by sensor noise) and the 0-dB bandwidth would vary from 54 Hz to 100 Hz with loop rates of 500 Hz to 1.8 kHz. Ultimately we discovered the reason for the higher relative bandwidth of the Robo-AO system was the readout time of the CCD39 was a much shorter 209 $\mu$s ($26\times26$ binned pixels read by 4 output amplifiers at an 806 kHz pixel rate). By comparison, the CCD60 used in Robo-AO-2 has a minimum readout time of 550 $\mu$s when used in the $2\times2$ binned mode which gives $4\times4$ pixels per subaperture on the laser wavefront sensor.

The path to increasing the correction bandwidth involves two areas of development: reducing the latency in the adaptive optics loop and in adding a proportional term to our control law. For the former, we have implemented a $4\times4$ binning mode with our wavefront sensor camera which reduces the readout time to 316 $\mu$s. Nüvü Camēras now offers cameras with a 30 MHz readout on these cameras (previously 20 MHz) and we had our camera modified with these new electronics, which has further reduced the readout time to 209 $\mu$s. We are also looking at ways to save time in our reconstruction software which has already benefited from parallel optimization\cite{software}. For the latter, we have implemented a proportional term to our loop control law. We have been able to test our implementation against our model and see consistency when using the internal telescope simulator. In that case we can easily push the 0-dB correction bandwidth to over 120 Hz without over amplifying errors at higher frequencies. On-sky, however, we have yet to see an increase in the correction bandwidth which we suspect is caused by the low SNR. We may not see a noticeable improvement until the laser signal photoreturn is increased. 

\subsection{Other improvements}

Currently we calibrate the slope-offset positions of the laser wavefront sensor using the internal telescope simulator UV light source. Because of the very small alignment errors in the telescope simulator, laser wavefront sensor, and, science arms, this should be adequate and static errors in the residual adaptive optics corrected science image shold me minimal. Despite this, significant static error remains as apparent in figs.~\ref{fig:iband}-\ref{fig:irdata2}. To address this, we will be using a version of the fast and furious focal plane wavefront sensing technique\cite{FF1, FF2} to update the slope-offset positions of the laser wavefront sensor. 

We will also be implementing the open-loop calibration of the deformable mirror (see Sec. 2.3). We have not expected this to make a considerable difference in the residual adaptive optics corrected wavefront because we are running the deformable mirror in a closed control loop, but it may help with the stability of the system, especially when using the proportional control on sky.  

\section{DISCUSSION}

The deployment of Robo-AO-2 is a crucial step to addressing many of the large survey and time domain science priorities as identified in 2020 United States Decadal Survey on Astronomy and Astrophysics\cite{roadmap}. Robo-AO-2 is currently able to provide diffraction limited imaging in red visible and near infrared wavelengths. Improvement in the delivered image quality is predicted over the next year as the UH2.2m is realuminized and we complete other optimizations. We also expect to start science campaigns once the new telescope control system is commissioned and automation of the facility is completed.

In late 2024 we anticipate the arrival of an adaptive secondary mirror for the UH2.2-m that uses very efficient hybrid variable reluctance actuators developed by TNO\cite{chun_asm1, chun_asm2}. The wavefront sensor and science camera suite on Robo-AO-2 will be used to evaluate the performance of this new technology on sky. 

\appendix    

\acknowledgments 

The Robo-AO-2 system is supported by the National Science Foundation under Grant No. AST-1712014, by the State of Hawaii Capital Improvement Projects, and by a gift from the Lumb Family. Support for the infrared camera for Robo-AO and Robo-AO-2 was provided by the Mt. Cuba Astronomical Foundation and through the National Science Foundation under Grant No. AST-1106391. Roboticization of the UH 2.2-m telescope is supported by the National Science Foundation under Grant No. AST-1920392.
 
The authors thank the people of the State of Hawaii for their support of astronomy. We are most fortunate to have the opportunity to work and conduct observations from public land on Maunakea. We are grateful to the UH2.2-m staff: Michael Yabe, Scott Caceres, Reid Ikeda and Maka Littorin; as well as Brian Landis, Charles Lockhart, Michael Kelii, Eric Warmbier, George Aukon, Richard Dekany, Michael Bottom and Michael Liu for their advice and help with Robo-AO-2. We thank Thomas Schneider for loaning us the reflectometer and for discussions on reflective coatings. 

\bibliography{Baranec_SPIE_2024} 
\bibliographystyle{spiebib} 

\end{document}